\documentclass[
twocolumn
,showpacs,preprintnumbers,amsmath,amssymb,prl,superscriptaddress]{revtex4}

\usepackage{amsmath}
\usepackage{dcolumn}
\usepackage{bm} 

\usepackage[dvipdfm]{graphicx}

\begin{document}
\title{Resonant X-Ray Diffraction Study of Strongly Spin-Orbit-Coupled Mott Insulator CaIrO$_3$}

\author{Kenya Ohgushi}
\affiliation{Institute for Solid State Physics, University of Tokyo, Kashiwa, Chiba 277-8581, Japan}
\affiliation{JST, TRIP, Chiyoda, Tokyo 102-0075, Japan}

\author{Jun-ichi Yamaura}
\affiliation{Institute for Solid State Physics, University of Tokyo, Kashiwa, Chiba 277-8581, Japan}

\author{Hiroyuki Ohsumi}
\affiliation{RIKEN SPring-8 Center, Sayo, Hyogo 679-8148, Japan}

\author{Kunihisa Sugimoto}
\affiliation{Japan Synchrotron Radiation Research Institute, SPring-8, Sayo, Hyogo 679-5198, Japan}

\author{Soshi Takeshita}
\affiliation{RIKEN SPring-8 Center, Sayo, Hyogo 679-8148, Japan}

\author{Akihisa Tokuda}
\affiliation{Department of Physics, Kwansei Gakuin University, Sanda, Hyogo 669-1337, Japan}

\author{Hidenori Takagi}
\affiliation{JST, TRIP, Chiyoda, Tokyo 102-0075, Japan}
\affiliation{Department of Advanced Materials Science, University of Tokyo, Kashiwa, Chiba 277-8561, Japan}
\affiliation{Advanced Science Institute, RIKEN, Wako, Saitama 351-0198, Japan}

\author{Masaki Takata}
\affiliation{RIKEN SPring-8 Center, Sayo, Hyogo 679-8148, Japan}
\affiliation{Japan Synchrotron Radiation Research Institute, SPring-8, Sayo, Hyogo 679-5198, Japan}

\author{Taka-hisa Arima}
\affiliation{JST, TRIP, Chiyoda, Tokyo 102-0075, Japan}
\affiliation{RIKEN SPring-8 Center, Sayo, Hyogo 679-8148, Japan}
\affiliation{Institute of Multidisciplinary Research for Advanced Materials, Tohoku University, Sendai, Miyagi 980-8577, Japan}

\date{\today}

\begin{abstract}
We performed resonant x-ray diffraction experiments at the $L$ absorption edges 
for the post-perovskite-type compound CaIrO$_{3}$ with $(t_{2g})^5$ electronic configuration.
By observing the magnetic  signals, we could clearly see that the magnetic 
structure was a striped order with an antiferromagnetic moment along the $c$-axis 
and that the wavefunction of a $t_{2g}$ hole is strongly spin-orbit entangled, the $J_{\rm eff} =1/2$ state.
The observed spin arrangement is consistent with 
theoretical work predicting a unique superexchange interaction in the 
$J_{\rm eff} =1/2$ state and points to the universal importance of the spin-orbit 
coupling in Ir oxides, irrespective of the local coordination and lattice topology. 
We also propose that the non-magnetic resonant scattering is 
a powerful tool for unraveling an orbital state even in a metallic iridate.
\end{abstract}

\maketitle

There is a new trend toward exploring Mott physics in a system with a 
strong spin-orbit interaction \cite{kim1, kim2, pesin, yang, watanabe}. 
Theoretical calculations on 
the Hubbard model revealed that the spin-orbit interaction drives 
a transition from a correlated metal to an insulator \cite{pesin, yang, watanabe}. 
This novel Mott insulating state is actually realized in a 
layered perovskite Sr$_{2}$IrO$_{4}$, including 
Ir$^{4+}$ ions with a $(t_{2g})^5$ electronic configuration \cite{kim1, kim2}. 
In this compound, one hole among $t_{2g}$ 
manifolds takes a complex wavefunction with the spin and orbital 
magnetic moments of $1/3$ and $2/3$ $\mu_{\rm B}$, respectively. 
This so-called $J_{\rm eff}=1/2$  state is achieved 
by the superiority of the spin-orbit interaction 
over the tetragonal crystal field around the $5d$ transition element. 

The superexchange interaction between two 
Ir$^{4+}$ ions in the $J_{eff}=1/2$ state is theoretically shown to be 
unique \cite{jackeli}. 
Whereas an antiferromagnetic Heisenberg 
interaction $J_{1}\bm{S_{i}} \cdot \bm{S_{j}}$ ($\bm{S_{j}}$ being the spin 
at the $j$-th Ir site) is dominant in a corner-shared 
IrO$_{6}$  bond (the Ir--O--Ir bond angle being 180$^{\circ}$), 
it completely vanishes in the edge-shared IrO$_{6}$ bond  
(the Ir--O--Ir bond angle being 90$^{\circ}$) owing to an interference effect. 
Instead, the magnetic interaction of the edge-shared bonds becomes a highly 
anisotropic and ferromagnetic one, $-J_{2}\bm{S_{i}^{z}} \bm{S_{j}^{z}}$, 
where the $z$ direction is perpendicular to the plane expanded by the two 
Ir atoms and two O atoms responsible for the edge-shared bond [Fig. 1(a)].
This interaction, which is called the quantum compass model, is 
unique in the sense that the spin anisotropy is produced by  
the perturbation effect of the Hund's coupling (not the spin-orbit interaction). 
To test the validity of this theory, it is necessary to elucidate 
the magnetic structure of an Ir oxide with an edge-sharing network. 


CaIrO$_{3}$ forms the post-perovskite structure shown in Fig. 1 (a). 
It is composed of edge-shared (corner-shared) IrO$_{6}$ octahedra along 
the $a$-($c$-)axis, where each octahedron is compressed along the corner-shared 
O direction (the $z$ direction) with a bond length ratio of 0.97 \cite{hirai}. 
This structure is in stark contrast to Sr$_{2}$IrO$_{4}$, where elongated 
IrO$_{6}$ octahedra with the bond length ratio 1.04 are  
connected solely by sharing corners. Hence, CaIrO$_{3}$ is an ideal platform for 
investigating the universal role of spin-orbit coupling in Ir oxides.
The compound shows a Mott insulating behavior characterized 
by the charge gap $\sim$0.17 eV and undergoes a transition to 
a possibly canted antiferromagnetic state at 115 K (= $T_{\rm N}$) \cite{ohgushi}.   
A deviation from the Curie-Weiss law above $T_{\rm N}$ as well as 
large coercive fields below $T_{\rm N}$ suggest an emergence of 
the spin-orbit interaction.

\begin{figure}[t]
\includegraphics[scale=0.25]{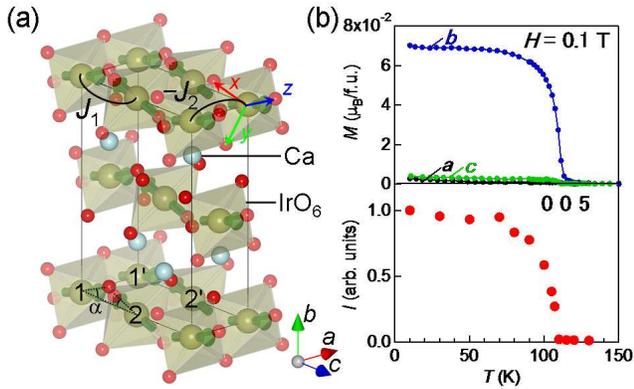}
\caption{(Color online) (a) Crystal structure of the post-perovskite CaIrO$_{3}$. 
The solid lines indicate the conventional unit cell, which is twice as large as 
the primitive unit cell. The magnetic interaction, as well as the magnetic structure 
determined in this study, are also shown. 
(b) Temperature ($T$) dependence of the magnetization ($M$) at 
the magnetic field ($H$) of 0.1 T \cite{supplementary} (upper), and the intensity ($I$) of 
the 0 0 5 reflection at $\psi = 0$ (lower). 
}
\end{figure}

In this Letter, we investigated the magnetic and orbital structure of CaIrO$_{3}$ 
by using the resonant x-ray diffraction \cite{blume1} at the $L$ absorption edges. 
This technique is particularly powerful in compounds with $5d$ transition metals 
because the wavelength of the x-ray beam is comparable to the lattice parameters 
\cite{kim2, mcmorrow, liu, ishii}
and the scattering amplitude is enhanced due to a dipole-allowed nature.
We also note that neutron scattering measurements are not applicable to
the present system with a strong neutron-absorbing element Ir.
We clarify that the magnetic structure of CaIrO$_{3}$ is 
a striped-type one and that the $J_{\rm eff}=1/2$ orbital state 
is realized also in a compressed octahedral coordination. 
We then compare these results with theoretical predictions.

Single crystalline CaIrO$_{3}$ was grown by 
the flux method. CaCO$_{3}$, IrO$_{2}$, and CaCl$_{2}$
with a molar ratio of 1:1:16 was slowly cooled 
from 1200 $^{\circ}$C to 1000 $^{\circ}$C for 240 h. 
Resonant x-ray diffraction measurements were performed at the beamline 
BL19LXU at SPring-8 \cite{supplementary, yabashi}. An incident beam was monochromated 
by a pair of Si (1 1 1) crystals and irradiated on the (0 0 1) surface of the sample, 
which was mounted in a $^{4}$He closed-cycle refrigerator installed on 
a four-circle diffractometer with a vertical scattering plane geometry. 
The intensities of incident and scattered beams were detected by an ionization chamber 
and a Si PIN photodiode, respectively. The polarization of the incident 
beam was perpendicular to the scattering plane $(\sigma)$ and that of the 
scattered beam was analyzed by using 
the 0 0 8 reflection of pyrolytic graphite. 
The azimuthal angle $\psi$ is defiled as $\psi =0^{\circ}$ when 
$\sigma$ $||$ $a$.
We also performed similar experiments at the beamline BL02B1 at SPring-8, 
where we use an imaging plate as a detector \cite{sugimoto}.

Figure 2 displays the absorption spectra obtained by fluorescence 
measurements at room temperature $(T)$ as well as 
the energy dependence of the scattered intensity of the 0 0 5 
reflection at $T=$ 10 K.  
At $\psi = 0^{\circ}$ [Fig. 2(b)], we can observe a strong 
resonance peak at the $L_{3}$ edge $\sim$11.2 keV;  
the intensity is about $0.015$ \% of the fundamental 
0 0 4 reflection. There are fine structures denoted by A and B 
with an integrated intensity ratio of 1:0.20, 
which origin will be addressed below. 
The space group of CaIrO$_{3}$ is $Cmcm$ orthorhombic symmetry, where 
0 0 $2n+1$ reflections are forbidden according to the $c$-glide reflection.
The polarization analysis 
indicates the $\pi'$ character of the scattered beam 
$I_{\sigma \mathchar`- \sigma'}/I_{\sigma \mathchar`- \pi'} = 3$ \% 
[inset of Fig. 2(b)], which also rules out the Thomson scattering as the origin.
The $T$ variation of the integrated intensity well  
follows that of the weak ferromagnetic moment [Fig. 1(b)].  
Considering also that the anisotropic tensor of susceptibility (ATS) 
scattering is prohibited in this geometry as discussed later, 
we conclude that the observed reflection originates from a commensurate 
antiferromagnetic order. Importantly, the 0 0 5 reflection 
cannot be detected within an experimental accuracy at the $L_{2}$ 
edge $\sim$12.82 keV, $I_{L_{2}}/I_{L_{3}} < 0.3$\%.  
At the $L_{3}$ edge, we also observed magnetic reflections at 
0 0 $l$ with $l = 1, 3, 7,$ and $9$.

The observed magnetic reflections at 0 0 $2n+1$ are well accounted for 
by considering an antiparallel arrangement of two Ir spins [labeled 1 and 2 in Fig. 1(a)] 
in a primitive unit cell.
In principle, one can also determine the spin direction experimentally 
by the $\psi$ dependence of the magnetic signal: 
$I_{\rm mag}\propto \sin^2\psi$, $\cos^2 \psi$, and $1$ for the spin direction 
along the $a, b,$ and $c$ axes, respectively \cite{magneticsignal}. 
However, the needle-like crystal morphology along the $a$ axis 
prevents us from performing such an analysis. We therefore employed the 
representation analysis, the results of which  are summarized in Table I. 
We note that the crystallographic space group for the magnetic phase below $T_{\rm N}$ 
has been revealed to be $Cmcm$ by oscillation photographs obtained at BL02B1.
Considering the parasitic ferromagnetism along the $b$-axis [Fig. 1(b)] 
together with the second order nature of the magnetic transition, 
we conclude that the $\Gamma_{3g}$ representation with the antiferromagnetic 
moments along the $c$-axis is realized. The obtained magnetic 
structure is schematically drawn in Fig. 1(a). It is a stripe-type 
order with a parallel alignment along the $a$-axis and 
an antiparallel alignment along the $c$-axis. 
This markedly contrasts with the checkerboard spin arrangement 
on the IrO$_{2}$ plane in Sr$_{2}$IrO$_{4}$ \cite{kim2}.

\begin{figure}[t]
\includegraphics[scale=0.7]{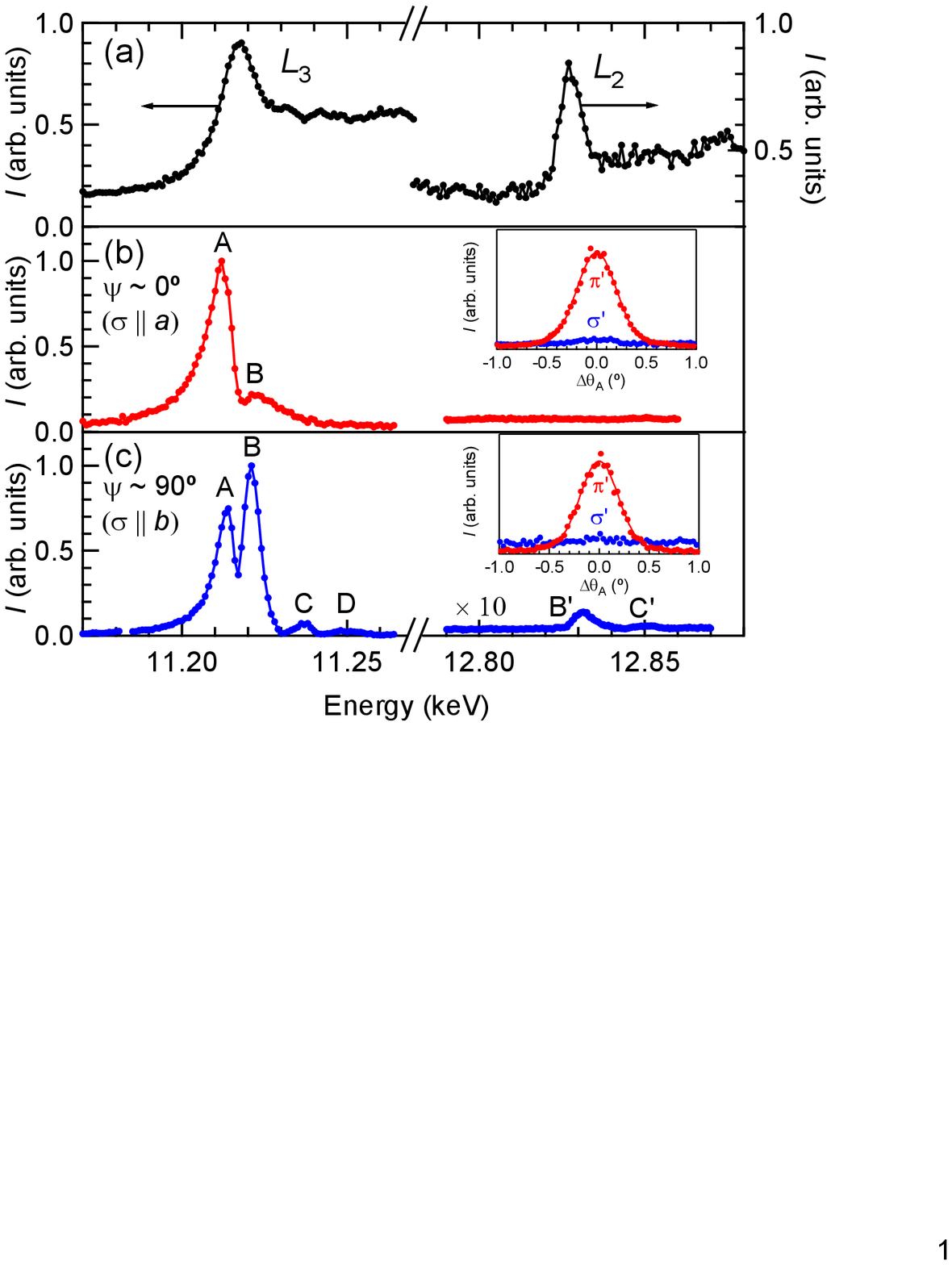}
\caption{(Color online) (a) X-ray absorption spectra near the Ir $L$ edge for CaIrO$_{3}$.   
(b) The energy dependence of the magnetic scattering intensity ($I$)
of the 0 0 5 reflection at $T=$ 10 K and $\psi =0^{\circ}$. 
The inset shows the polarization analysis of the scattered light 
at the $L_{3}$ edge, where $ \Delta \theta_{\rm A}$ represents the analyzer angle.  
(c) Same data as (b) except for $\psi =90^{\circ}$. The ATS scattering 
is dominant in this geometry. Note that 
the data at the $L_{2}$ edge is magnified by 10 for clarity.
}
\end{figure}

\begin{table}[b]
\caption{The magnetic structures of the irreducible representations ($\Gamma_{i}$) 
for the $Cmcm$ space group. $M$ denotes the magnetic point group, where 
$\underline{1}$ stands for the time-reversal operation.
$a, b, $ and $c$ represent the spin directions.
AF and F denote the antiferromagnetic and ferromagnetic arrangement of spins at 
two Ir sites in a unit cell. } 
\begin{ruledtabular}
\begin{tabular}{cccccc}
$\Gamma_{i}$  & $M$ & generators  & $a$ & $b$ & $c$ \\
\hline
$\Gamma_{1g}$ & mmm & $2_{x}$, $2_{y}$, $\overline{1}$
& AF & $\cdot$ & $\cdot$ \\
\hline
$\Gamma_{2g}$ & m$\underline{{\rm mm}}$ & $2_{x}$, $\underline{2_{y}}$, $\overline{1}$
& F & $\cdot$ & $\cdot$ \\
\hline
$\Gamma_{3g}$ & $\underline{{\rm m}}$m$\underline{{\rm m}}$ & $\underline{2_{x}}$, $2_{y}$, $\overline{1}$
&$\cdot$ & F & AF \\
\hline
$\Gamma_{4g}$ & $\underline{{\rm mm}}$m & $\underline{2_{x}}$, $\underline{2_{y}}$, $\overline{1}$
& $\cdot$ & AF & F \\
\end{tabular}
\end{ruledtabular}
\end{table}

We move to the wavefunction of a $t_{2g}$ hole.
When the tetragonal crystal field $\Delta$ ($> 0$ for the 
compressed octahedron) and the spin-orbit coupling $\zeta$ are present,
the sixfold degenerated $t_{2g}$ orbitals are split into 
three doubly degenerated bands.
At the ground state, one hole occupies one of the 
highest energy orbitals $|\varphi, \pm> = 
\frac{1}{\sqrt{A^2+2}} (A |xy, \pm>\pm|yz, \mp>+i|zx, \mp>)$,
where $A = \frac{-\zeta -6 \Delta +\sqrt{3}\sqrt{3 
\zeta^2+4 \zeta \Delta +12 \Delta^2}}{2 \zeta}$ quantifies 
the role of the tetragonal crystal field: $A=0$ when 
$\Delta/\zeta = \infty$ and $A=1$ when $\Delta/\zeta = 0$; 
the latter corresponds to the $J_{\rm eff}=1/2$ state.
There has been a simple argument that the orbital character can be 
determined by the edge dependence of the magnetic scattering intensity \cite{kim2}. 
We here discuss this issue in more detail on the basis of 
CaIrO$_{3}$ \cite{supplementary}.

When one electron is virtually excited from the $2p$ orbitals 
in a resonant process, the $t_{2g}$ bands are fully occupied; 
this simplicity enables us to calculate 
the atomic scattering tensor straightforwardly. 
By combining contributions of two Ir sites in a 
striped-ordered state, we obtain the tensor 
structure amplitude $\hat{F}$ for the 0 0 $2n+1$ reflection. 
Nonzero components in $\hat{F}$ are 
$F_{ab} = -F_{ba} \equiv i F_{\rm mag}$ and $F_{bc} = F_{cb} \equiv F_{\rm ATS}$ \cite{dmitrienko}. 
The $F_{\rm mag}$ term changes its sign by applying the time-reversal operation, 
and represents the magnetic scattering. 
The $F_{\rm ATS}$ term corresponds to the ATS scattering. 
The local principal axes towards the corner-shared O, $z$, of two Ir sites 
in a primitive unit cell are alternately 
titled by an angle $\pm \, \alpha (\sim$23$^{\circ})$ from the $c$-axis [Fig. 1(a)]. 
This difference in the local anisotropy revives 0 0 $2n+1$ reflections, which 
are pronounced in a resonant condition.
The scattering intensity shows the polarization and azimuthal dependence as follows:  
$I_{\sigma \mathchar`- \sigma'} = 0$ and 
$I_{\sigma \mathchar`- \pi'} = | \cos \theta \sin \psi F_{\rm ATS}+i  
\sin \theta F_{\rm mag}|^2$, where $\theta$ is the Bragg angle.
The scattering intensity also shows a large difference between at 
the $L_{2}$ and $L_{3}$ edges. 
We can quantitatively estimate this by referring to 
calculated results: $F_{\rm mag}=\pm (\cos \alpha/3)f(A^2-2A-2)/(A^2+2)$ and 
$F_{\rm ATS}=-(\sin 2\alpha/3) f(A^2+A-2)/(A^2+2)$ for the $L_{3}$ edge; and 
$F_{\rm mag}=\mp (\cos \alpha/3) f(A-1)^2/(A^2+2)$ and 
$F_{\rm ATS}=-(\sin 2\alpha/6) f(A-1)^2/(A^2+2)$ for the $L_{2}$ edge.
The intensity ratio of the $L_{2}$ to $L_{3}$ edges are plotted in Fig. 3.
One can see that both the magnetic and ATS scattering intensity ratio 
steeply decrease with approaching to the $J_{\rm eff}=1/2$ state ($A=1$) 
and that particularly the ATS scattering ratio with a larger value 
is more useful to determine the orbital character in a large $A$ 
region than the magnetic scattering ratio.

We now focus on the spectra at $\psi =90^{\circ}$ [Fig. 2(c)], 
where the ATS scattering becomes largest.
The $\psi$-independent magnetic scattering is overlapping; however, 
we can conclude that the ATS scattering is dominant at $\psi =90^{\circ}$, 
$I_{\rm ATS}/I_{\rm mag} \sim 30$ at the $L_{3}$ edge, by 
comparing the intensity at $\psi =0$ and $90^{\circ}$. 
We confirmed a $\pi'$ character of the scattered beam
$I_{\sigma \mathchar`- \sigma'}/I_{\sigma \mathchar`- \pi'} =3.9$\% 
[inset of Fig. 2(c)].

At the $L_{3}$ edge, we can discern four fine structures centered at 
11.214, 11.221, 11.237, and 11.249 keV with an integrated intensity 
ratio of 1:0.95:0.06:0.04. The first and second peaks, the energies of which
coincide with inflection points of the absorption spectrum, 
are also observed at $\psi = 0^{\circ}$ and are ascribed to  
the virtual excitation from the $2p$ $J =3/2$ orbitals 
to the $t_{2g}$ and $e_{g}$ orbitals, respectively (inset of Fig. 3).
The energy difference of these two peaks corresponds to 
the crystal field splitting due to the octahedral crystal field,    
which is estimated to be $10Dq = 7$ eV. This value is much larger 
than typical values in the $3d$ and $4d$ transition metal oxides, 
e.g. $10Dq = 4$ eV in Ca$_{2}$RuO$_{4}$ \cite{zegkinoglou}. 
This can be interpreted as the strong covalency effect in a 5$d$ 
transition metal enlarging the band splitting.
The higher energy structures denoted by C and D are likely related 
to the $6s$ and $6p$ bands.

At the $L_{2}$ edge, as in the case of $\psi =0^{\circ}$, 
we could not observe any signal at the energy related to $t_{2g}$ orbitals
as an intermediate state. We note that a peak structure centered 
at 12.832 keV corresponds to an inflection point of the 
absorption spectrum at the {\it higher} energy side; hence, 
the intermediate state is $e_{g}$ orbitals. 
The intensity ratio at the $t_{2g}$ related energy is 
$I_{L_{2}}/I_{L_{3}} < 0.05$\%. 
Comparing this observation with calculated results shown in Fig. 3
imposes a restriction of $A> 0.87$. On the other hands, the finite ATS 
signal at the $L_{3}$ edge indicates $A< 1$. We thus conclude that 
a $t_{2g}$ hole has a slightly modified $J_{\rm eff} =1/2$ state.
We here point out that the ATS scattering intensity is a
quite sensitive probe of the wave function. 
It should be interesting to apply this probe to metallic 
iridates with no magnetic order \cite{moon, korneta, ohgushi}, 
where various exotic quantum states are anticipated \cite{yang, wang}.

\begin{figure}[t]
\includegraphics[scale=0.55]{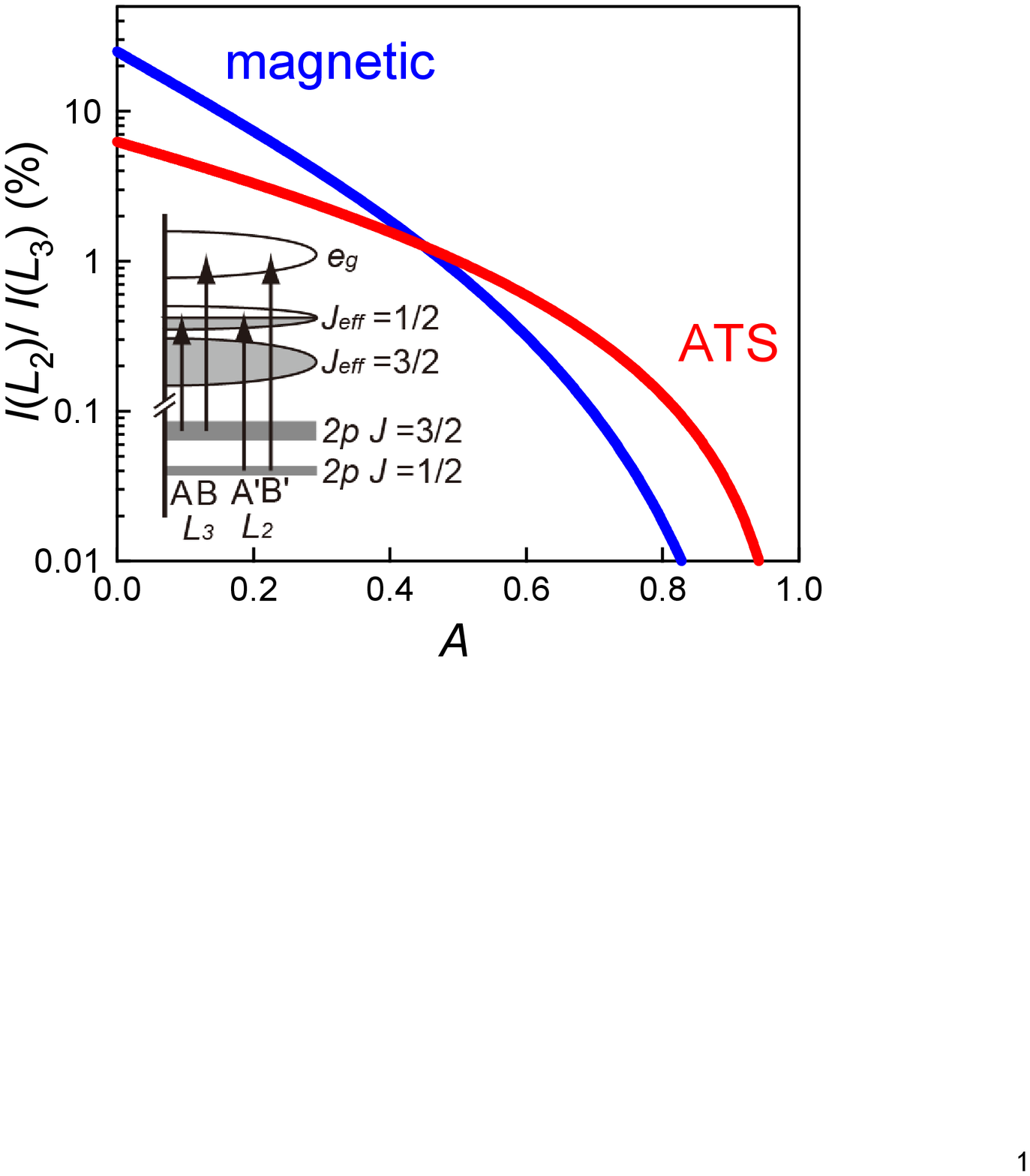}
\caption{(Color online)  
The calculated intensity ratio in the magnetic and ATS scattering 
of the $L_{2}$ to $L_{3}$ edges as a function of 
the $A$ coefficient. The $A =1$ corresponds to the 
$J_{\rm eff}=1/2$ state. Note that the ordinate is scaled logarithmically.
The inset shows the energy scheme 
of the virtual excitations in resonant processes.    
}
\end{figure}

Our result that the $J_{\rm eff} =1/2$ state is realized in 
CaIrO$_{3}$ enables us to argue the magnetic structure   
in the framework of the theory by Jackeli and Khaliullin unambiguously \cite{jackeli}. 
We recall that the theory predicts an antiferromagnetic (ferromagnetic) 
interaction through the corner-(edge-)shared bonds, 
being consistent with our spin arrangements. 
Moreover, the weak ferromagnetism along the $b$-axis 
can be successfully explained by the theory. 
There are two mechanisms which induce the spin canting.
One is that the anisotropic axis $z$ in the quantum compass model is 
distinguishable between the Ir(1)--Ir(1') and Ir(2)--Ir(2') 
bonds [Fig. 1(a)]. Another is the Dzyaloshinskii--Moriya (D--M) 
interaction at the Ir(1)--Ir(2) bond with the D--M vector $\bm{D}_{1,2}=(D, 0, 0)$
(the site symmetry at the midpoint of the bond being $m2m$).
A mean field treatment of the Hamiltonian including $J_{1}$- and $J_{2}$-terms, and 
the D--M interaction gives the most stable spin arrangement to be the 
experimentally observed stripe-type order with the spin canted angle of $\alpha$.
On the other hands, when we assume the antiferromagnetic magnetic 
moment of 1 $\mu_{\rm B}$/Ir, which is the expected value for the 
completely localized $J_{\rm eff}=1/2$ state, the observed uniform magnetic moment 0.07 
$\mu_{\rm B}$/Ir indicates the canted angle $\sim$4$^{\circ}$. 
This is much smaller than $\alpha \sim$23$^{\circ}$.  
The reason for this discrepancy between the theory and experiment 
is likely related to a reduced spin moment due to quantum fluctuations, 
a slight deviation from the $J_{\rm eff}=1/2$ state, 
or the effect of the direct interaction 
which is not considered in the present analysis.

To summarize, we have investigated the spin and orbital state of CaIrO$_{3}$  
post-perovskite by the resonant x-ray diffraction. The orbital of a 
$t_{2g}$ hole is the  $J_{\rm eff} =1/2$ state. This strongly spin-orbit 
coupled state stabilizes a striped-type magnetic order.
We also propose the anisotropic tensor of susceptibility scattering 
as a tool to investigate the orbital character in a metallic iridate.

We thank Y. Ueda for the use of a single crystalline x-ray diffractometer.  
The synchrotron radiation experiments were performed at BL19LXU and BL02B1
in SPring-8 with the approval of RIKEN (Proposal No. 20090064) and the Japan 
Synchrotron Radiation Research Institute (JASRI) (Proposal No. 2009B1200), respectively.
This work was supported by Special Coordination Funds for Promoting 
Science and Technology, Promotion of Environmental Improvement for Independence 
of Young Researchers, and Grant-in-Aid for Scientific Research (B) (No. 20740211).


\begin{thebibliography}{}
\bibitem{magneticsignal} 
The magnetic reflection can be written as 
$I_{\rm mag} \propto |\sum_j \hat{R} \bm{S}_{j} \cdot \bm{\epsilon}_{i} 
\times \bm{\epsilon}_{s}$e$^{2 \pi i \bm{q} \cdot \bm{r}_{j}} |^2$,  
where $\bm{q}$ is the momentum transfer and $\bm{r}_{j}$ is 
the position of the $j$-th Ir atom.
\end{thebibliography}
\end{document}